# Probing the Spin-Momentum Locking in Rashba Surfaces via Spin Current

*José E. Abrão[1,*], Eudes Gomes da Silva[1,3], Gilberto Rodrigues-Junior[2], Joaquim B. S. Mendes[2,*], Antonio Azevedo[1,*]*

[1]Departamento de Física, Universidade Federal de Pernambuco, 50670-901 Recife, PE, Brazil.
[2]Departamento de Física, Universidade Federal de Viçosa, 36570-900 Viçosa, MG, Brazil.
[3]Department of Physics and Astronomy, University of Iowa, Iowa City, Iowa, USA.

In this work, we explore the intriguing spin-momentum locking phenomenon within the Rashba states of antimony (Sb) films. By combining spin pumping with the flow of an external charge current, we reveal the topological properties of surface states in Sb films. Taking advantage of the well-defined spin polarization of both spin-momentum-locked charge currents and spin-pumped currents, we demonstrate precise manipulation over the direction and magnitude of the resulting charging current, generated through the inverse Rashba-Edelstein effect. This fascinating phenomenon is attributed to the dynamic interaction between the accumulation of out-of-equilibrium pumped spins and the flowing spins, intrinsically locked perpendicular to the direction of the charge current. The results show that Sb as a promising material for basic and applied investigation of spintronics phenomena. We believe that the nanostructures investigated here open the way for the development of low-power logic gates operating in the range of a few tens of microamperes.

- **INTRODUCTION**

In the fascinating realm of condensed matter physics, spintronics has emerged as a groundbreaking field that exploits the spin of electrons for information processing and storage. Unlike traditional electronics, which relies solely on the charge of electrons, spintronics harnesses both the charge and spin of electrons, providing a new pathway for creating more efficient and versatile electronic devices. [1,2,3] At the core of spintronics lies the fundamental concept of spin currents - currents that represent the flow of spin angular momentum. These currents can be used to transport information, to exert torques on magnetic materials, as well as converting the spin current into a change current and vice-versa. [4]

One prominent mechanism used to describe the spin-to-charge conversion in bulk materials is the direct and inverse spin Hall Effect. [5] The direct Spin Hall effect (SHE) is a phenomenon where an applied electric field induces the generation of a pure spin current perpendicular to its direction. Unlike the conventional Hall Effect, which involves the separation of charge carriers based on their charge, the direct Spin Hall effect separates the charge carriers based on their spin states. The SHE is a consequence of spin-orbit coupling, an interaction between the spin of electrons and their motion in a crystal lattice with structural asymmetry [5] or with a scattering center such as impurities or crystal defects. [6,7] Conversely, a reciprocal effect entitled the inverse Spin Hall effect (ISHE) exists. The ISHE is a phenomenon where a pure spin current, injected into a material exhibiting strong spin-orbit coupling, induces a charge accumulation perpendicular to both the spin polarization $\hat{\sigma}$ and the injected spin current direction $\vec{J}_S$, the charge current generated by the ISHE is described by:

$$\vec{J}_c = \theta_{SH}\left(\frac{2e}{\hbar}\right)(\hat{\sigma} \times \vec{J}_S). \qquad (1)$$

Where $\theta_{SH}$ is the Spin-Hall angle that represents the efficiency of the spin-to-charge conversion process, $\hbar$ denotes the reduced Planck constant and $e$ signifies the elementary charge. The electrical voltage arising from ISHE-induced charge accumulation provides an experimental manner to translate spin information into a conventional electrical signal. As a fundamental mechanism in spintronic, ISHE holds promise for the development of memory, logic, and sensing devices. Moreover, it serves as a valuable tool for probing the physical properties of materials. [8,9]

While SHE and ISHE are appropriate for bulk materials, surfaces and interfaces require consideration of additional significant effects: the direct and inverse Rashba-Edelstein effects. The direct Rashba-Edelstein effect (REE) emerges when electrons move through a crystal lattice without inversion symmetry, while simultaneously exhibits spin-orbit coupling. This effect can be understood as the manifestation of an effective magnetic field, even in the absence of an external magnetic field, on electrons in motion in their own reference frame. [10,11] The effective magnetic field then couples to electron's magnetic moment through a Zeeman-like energy $\mu_B \vec{\sigma} \cdot \vec{B}$, where $\vec{\sigma}$ and $\mu_B$ represent the vector of Pauli matrices and the Bohr magneton, respectively. Thereby, producing a steady non-equilibrium spin polarization with opposite sign on opposite edges of the sample. [12] The hallmark of the Rashba-Edelstein effect is the formation of Rashba spin-orbit splitting, where electrons with opposite spins experience different energies and momenta, causing them to move in opposite directions, this property is known as spin-momentum locking and it is represented in Fig. 1(a). For surfaces and interfaces the inversion symmetry is broken due to absence/change of neighboring atoms which produces an electric field as it is represented in Fig. 1(b). It is worth mentioning that in recent works on effects in 2D gases with strong spin-orbit coupling, the theory of the Rashba-Edelstein Effect is developed based on quantum kinetic equations and diagrammatic calculations for systems in which the spin-orbit interaction is linear in $\vec{k}$. [13,14,15]

Like the SHE/ISHE effects, a reciprocal effect of the direct Rashba-Edelstein effect, namely the Inverse Rashba-Edelstein Effect, exists. In the inverse effect a non-equilibrium spin density $\vec{S}_{neq}$, which is normally induced into the system by the injection of a pure spin current, induces a charge accumulation perpendicular to both $\vec{S}_{neq}$ and the direction $\hat{z}$ of symmetry breaking field, the charge current generated by the ISHE is described by: [16,17]

$$\vec{J}_C = \frac{e\alpha_R}{\hbar}(\hat{z} \times \vec{S}_{neq}). \qquad (2)$$

A hallmark of the inverse Rashba-Edelstein effect is that the charge current converted by this effect does not depend on the direction of the spin-current injected into the system, it depends only on the spin polarization $\hat{\sigma}$ which dictates the direction of $\vec{S}_{neq}$.

The direct and inverse Rashba-Edelstein Effect has being explored in different systems such as graphene, transition metal dichalcogenides (TMDs), LaAlO$_3$/SrTiO$_3$ and Bi/Ag interfaces, and in topological insulators.[18-25] Among materials with surface states, antimony (Sb) has proven to be a unique material, not only it is an elemental material, Sb is a constituent element of several topological materials and can be a topological insulator by itself. [26,27] Furthermore, a recent study showed that the Sb films can be grown by sputtering without the need of heat treatment or the use of special substrates and still presents surface states. [28] Thus, Sb is a material that presents unique possibilities for different experiments and applications.

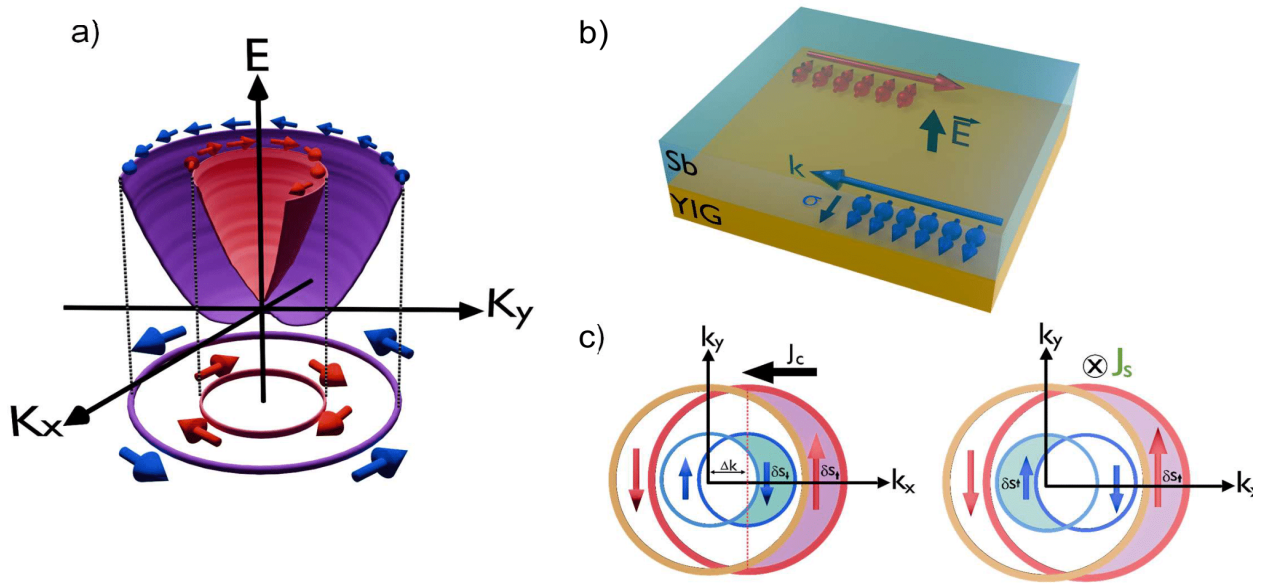

**Figure 1.** Schematics of the Rashba-Edelstein Effects: in (a) is shown the energy splitting due to the Rashba spin-orbit coupling, moreover it is possible to observe the spin-momentum locking, where electrons with different spins travel through opposite directions. In (b) it is shown a schematic representation of the spin-momentum locking, characteristic of surface states, in the YIG/Sb layer and in (c) it is presented the direct and inverse Rashba-Edelstein effects in a 2D electron gas. In the Direct REE, an electric field exerts a force on the electrons resulting in a displacement of the Fermi circles, this displacement produces a non-equilibrium spin accumulation in a perpendicular direction. The Inverse REE, a spin current pumped into the interface creates a spin accumulation that induces the displacement of the Fermi circles resulting in a charge current in a perpendicular direction.

In a recent work, we explored the spin-to-charge conversion phenomenon in Co/Sb/Py heterostructure. [28] Our findings revealed that the spin-to-charge conversion in Sb thin films was not dependent on the direction of the injected spin current, even for thicker films. Thus, no ISHE was observed in this material. Instead, the measured signal could only be attributed to IREE occurring at the Co/Sb and Sb/Py interfaces. To further explore antimony properties, in this current work we perform spin pumping measurements, while applying an external charge current. The samples consisted of YIG/Sb(15nm)/Py(12nm) and YIG/Sb(15nm)/Ti(3nm) heterostructures, where YIG and Py stand for yttrium iron garnet ($Y_3Fe_5O_{12}$) and permalloy ($Ni_{81}Fe_{19}$), respectively. The YIG films were produced by means of the Liquid Phase Epitaxy (LPE) following the traditional $PbO/B_2O_3$ flux method, [29] while the Sb, Ti and Py films were deposited by DC-Sputtering. The structural properties, surface roughness and morphology of sputtered grown Sb thin film are depicted in Figures S1-S3 of the supplementary material in this manuscript. X-ray diffraction patterns confirm the crystalline and low surface roughness of the films, with preferential growth along the c-axis (see the Supplemental Information for more details about the conditions of growth and the crystallographic structure of the Sb films).

### ▪ RESULTS AND DISCUSSION

The spin pumping driven by ferromagnetic resonance (SP- FMR) measurements were made via a ferromagnetic resonance (FMR) spectrometer at room temperature, operating at 9.42 GHz, and all the investigated samples were cut into sizes of 2 x 3 $mm^2$. We mounted the sample on a PVC rod and placed it at the bottom of a rectangular microwave resonant cavity ($TE_{102}$ mode), where the RF magnetic field is maximum, and the RF electric field is minimum. Four electrodes arranged in a four-point configuration (see Fig. 2(a)), were then attached to the sample with silver paint. This configuration enables the detection of the spin pumping signal while also being able to apply a charge current to the sample. As YIG is an insulator, the flow of charge current will pass only through the Sb, thus ensuring that any changes in the spin pumping signal due to the current applied to the film are exclusively attributed to the surface states.

The main objective of the initial experimental verification was to determine whether Sb, when grown onto YIG, maintains its surface states. To achieve this, spin pumping measurements were performed on YIG/Sb/Py trilayers with no applied current. If the system truly exhibits surface states and the spin to charge conversion is dominated by IREE, the spin pumping signal will not depend on the spin current direction. In this scenario, only the polarization of the spin current, which can be fixed by the external field, becomes significant. Consequently, the observation of two peaks with the same polarity is expected.[16] Fig. 2 shows both the ferromagnetic resonance (Fig. 2(c)) and the spin pumping signal for the YIG/Sb(15nm)/Py(12nm) trilayer (Fig. 2(b)) with no current passing through the Sb layer. It is noteworthy that both signals are positive, indicating that the antimony grown on top of YIG indeed possess surface states. In fact, the signature of topologically surface states on sputtered grown Sb thin film can be corroborated through scanning tunneling spectroscopy measurements, as represented in Fig. S4 of the supplementary information.

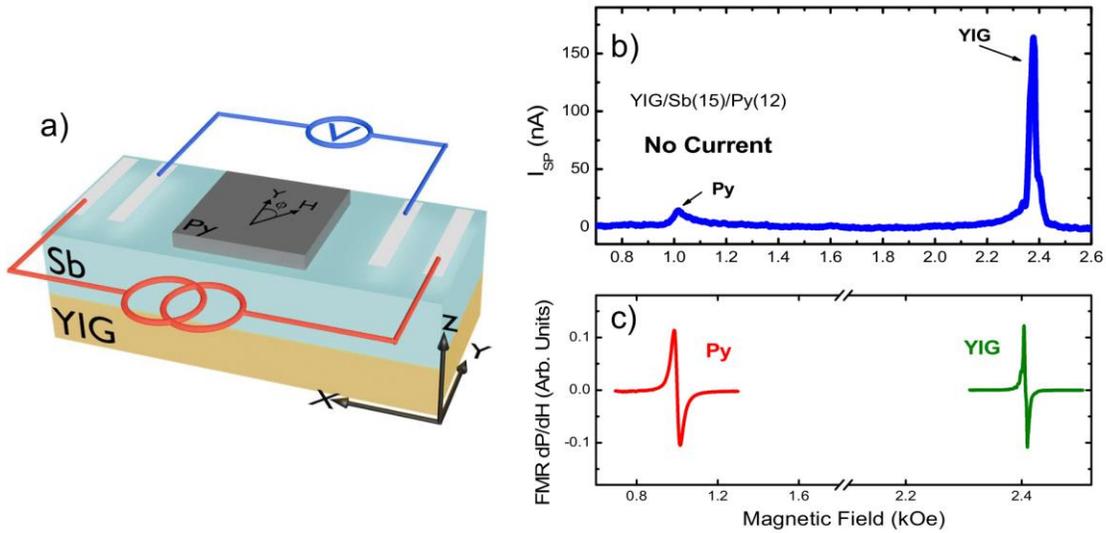

**Figure 2.** (a) Schematic of the experimental setup used for simultaneous measurements of spin pumping and DC current application, where electrodes were attached to the edges of the sample in a four-point configuration. (b) Spin pumping signal for YIG/Sb(15nm)/Py(12nm), with both signals displaying the same polarity. The applied RF power was 110 mW. (c) Derivative of the absorption FMR signal for both ferromagnets, presented separately due to the difference in microwave power needed to excite the FMR condition. To excite the FMR in Py and YIG we used 110 mW and 0.01 mW, respectively.

Furthermore, a subtle asymmetry is observed in the signal corresponding to the YIG resonance in Figures 2(b, c). This asymmetry indicates that an excessive use of RF power to drive the FMR condition, leading to the onset of nonlinear effects in the YIG. To mitigate these effects, it is necessary to reduce the microwave power used to excite the resonance. However, decreasing the RF power causes the Py absorption signal to not be detected on the same scale as the YIG signal. Consequently, we decided to disregard the Py signal and focus exclusively on the YIG resonance.

After confirming that antimony grown onto YIG also exhibits surface states, measurements were conducted with an applied DC current. Initially, FMR measurements were performed with and without the external current, as shown in Fig. 3(a). Notably, no significant changes were observed, with the FMR resonance field and linewidth remaining constant at $H_R = 2.415$ kOe and $\Delta H = 2.4$ Oe, respectively, across all resonance spectra. Due to the exceptional quality of our YIG film, magnetostatic modes are excited both below (surface modes) and above (volume modes) the FMR field. This indicates that the current flowing through the Sb layers does not introduce extra damping or change the FMR field. The spin pumping measurements presented in Fig. 3(b) showed very different results. Thus, without any applied current, the signal for $\phi = 180º$ exhibits a single, well-defined peak pointing upwards. Upon applying a current of -1.0 mA to the antimony layer, the measured signal increased by almost fourfold. Interestingly, when inverting the applied current to +1.0 mA, not only does the signal become larger, but it also changes polarity. Moreover, numerical curve fitting revealed that the peak signal obtained for an applied current of -1.0 mA ($I_A$) added to the signal for +1.0 mA ($I_B$) is equivalent to the spin pumping signal without an external current ($I_c$). To gain deeper insights into the impact of the applied current, a series of measurements were conducted by varying its magnitude and polarity. The peak signal measured in the spin pumping experiment showed a linear behavior with the applied current, as shown in Fig. 3(c).

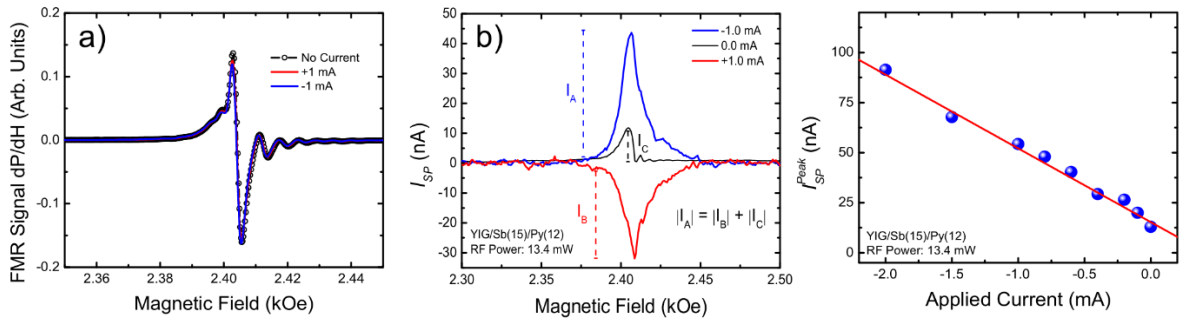

**Figure 3.** (a) FMR absorption curves for a typical a YIG/Sb(15nm)/Py(12nm) sample with ± 1.0 mA and no applied current, confirming that in all cases there is no change in the FMR field and linewidth. (b) Comparison of the spin pumping signal for ± 1.0 mA and no applied current, highlighting the signal change polarities with respect to the applied current. (c) Linear dependency of the peak of the spin pumping signal as function of the applied current.

The sample studied so far had a top layer of Py as there was initially an interest in verifying whether antimony grown under YIG presented surface states. However, one might question whether Py influences the spin pumping measurement previously presented with applied current due to possible contamination of the measured signal via galvanomagnetic effects.[30,31] To resolve this concern, YIG/Sb(15nm)/Ti(3nm) samples were prepared, where the Ti layer completely covers the Sb layer to protect it from natural oxidation. Fig. 4 shows the results obtained by applying a current to the YIG/Sb(15nm)/Ti(3nm) sample. Notably, the results are similar to those previously observed. The application of current changes the measured signal drastically, being enhanced by more than four times just by applying a current of -1.0 mA. Moreover, changing the polarity of the applied current from -1.0 mA to +1.0 mA the signal also changes its polarity, as observed in Fig. 4(b). Based on these results, we can certainly conclude that the Py layer did not exert any influence on the system; instead, it merely served as a protective barrier to prevent oxidation. Moreover, it was also observed that the measured SP-FMR signal exhibited a linear relationship with the applied current as shown in Fig. 4(c), where the insert shows the peak value as function of the applied current. The linear fit indicates that a current of -41.7 µA is enough to compensate for the spin pumping signal. This result is of interest as it shows that it is possible to build low powered logic devices with Rashba-interfaces, where both positive and negative spin pumping signals can be used as distinct logic levels. To further understand the system under study, measurements varying the microwave power while keeping the external applied current fixed at -0.2mA were performed, and the results in Fig. 4(d) show a comparison of the spin pumping signal for different values of microwave power used to promote the FMR condition. The numerical fit of each curve reveals that the peak signal behaves in a linear manner with the microwave power as it is presented in the insert of Figure 4(d).

The intriguing findings we observe can be explained by the interplay between the accumulation of non-equilibrium spins in Rashba states, arising from spin pumping and spin-momentum locking phenomena. In the REE, the passage of an electrical current trough a Rashba interface separates the electrons based on their spin states, which naturally give rises to a non-equilibrium spin density due to the difference in energy of each spin band (See Fig 1(a) and 1(c)). In the spin pumping process, the magnetization dynamics produces a spin accumulation at the YIG/Sb interface. This accumulation depends on the direction of the DC magnetic field and the RF power, and leads a non-equilibrium spin density by shifting one of the Rashba surfaces at the Fermi level (See Fig 1(c)). Thus, introducing spin accumulation on a Rashba surface results in the emergence of a spatially uniform out-of-equilibrium spin density. This is manifested by changes in the Fermi surface radius for each spin state. [12,20] Fig. 1(c) illustrates the direct and inverse Rashba-Edelstein effects, where the non-equilibrium spin density is illustrated by the difference in the colored areas representing up and down spin densities.

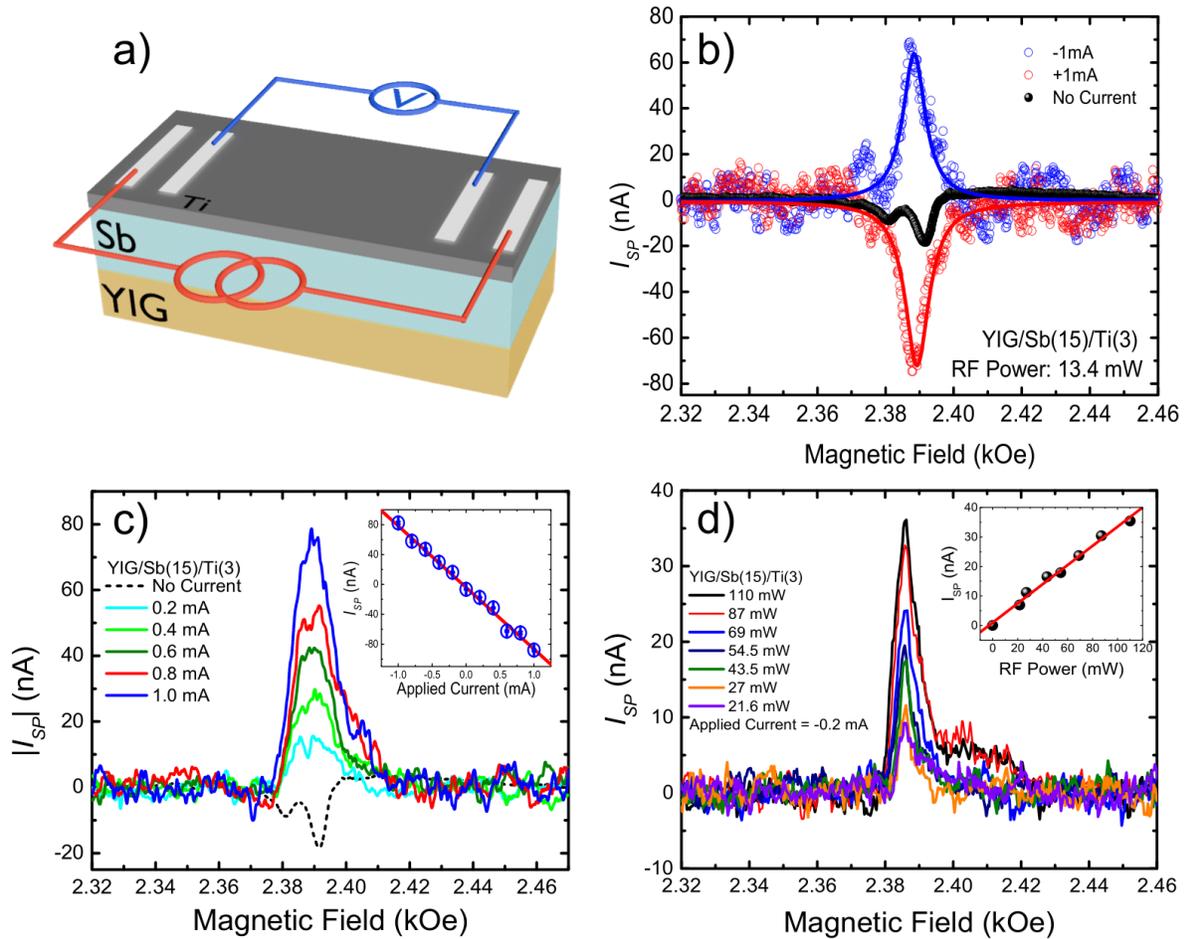

**Figure 4.** (a) Schematic illustration describing the measurement setup of the SP-FMR for a typical YIG/Sb(15nm)/Ti(3nm) heterostructure. (b) Presentation of the spin pumping signal for ± 1.0 mA of applied current and the signal with no applied current. Note that the signal reverses with changes in the direction of the applied current. (c) Comparative analysis of the spin pumping signal for positive values of applied current. The insert shows the linear dependence of the peak spin pumping signal as function of the applied current. (d) Comparative analysis of the spin pumping signal as a function of microwave power used to promote the FMR condition with an applied current of -0.2 mA passing through the sample. In the insert it is presented the linear dependence of the measured signal as function of the electromagnetic radiation power.

If we consider a scenario in which a charge current $\vec{J}_C$ is applied along the x-axis, it will result in a non-equilibrium spin density along the y-axis due to REE, denoted as $\vec{S}_{neq}^C$. At the same time, the spin pumping technique also induces a non-equilibrium spin density, represented by $\vec{S}_{neq}^{SP}$, at the interface. However, its direction depends on the direction of the external applied field, which pins the polarization of the spin current injected into the YIG/Sb interface. Depending on the relative orientation between $\vec{S}_{neq}^C$ and $\vec{S}_{neq}^{SP}$, the resulting spin density increases (when $\vec{S}_{neq}^C$ and $\vec{S}_{neq}^{SP}$ are parallel) or decreases (when $\vec{S}_{neq}^C$ and $\vec{S}_{neq}^{SP}$ are antiparallel). Thus, in the conducted experiments, there are two sources contributing to the resultant out-of-equilibrium spin density,

$$\vec{S}_{neq} = \vec{S}_{neq}^{SP} + \vec{S}_{neq}^C. \qquad (3)$$

The first contribution arises from the non-equilibrium spin polarization due to the injection of a spin current by SP-FMR into the Rashba states. In our experiment, this contribution depends on the experimental parameters related to the FMR, mainly the rf power. By varying the rf power we observed a linear dependency which is expected by due to the spin pumping effect (see Fig. 4(d)). The second term contributing to the out-of-equilibrium spin density comes from the electric current applied to the system via the direct REE. This term varies with changes in the applied current, and notably, its spin polarization can invert with a reversal in the applied current. The results presented in figures 3 and 4 can be explained solely by inverse Rashba-Edelstein effect in Eq. (2) if one considers that $\vec{S}_{neq}$ contains a component dependent on the applied current passing through the Sb layer.

To validate the hypothesis regarding the two terms that make up the non-equilibrium spin density, an experimental setup was developed in which now the current is applied perpendicular to the direction where we measured the spin pumping signal, represented in the insert of Fig. 5(a), similar to a planar Hall effect setup. In this new configuration, it is important to note that the out-of-equilibrium spin density due to REE is perpendicular to the direction of the applied current. Thus, when a current is applied in the $\hat{y}$ direction, the resulting out-of-equilibrium spin density is oriented along the $\hat{x}$ direction. Consequently, it is anticipated that the electric signal produced by the spin-momentum locking effect, is not detected due to the chosen measurement geometry. Fig. 5(a) shows a comparison of the spin-pumping signal, measured along $\hat{x}$ direction, with applied currents of $\pm 1.0$ mA along $\hat{y}$ direction. Notably, the SP-FMR signal does not change, meaning that the addition of current to the system only leads to an increase in the measurement noise, likely attributed to the thermoelectric nature of Sb. These results are consistent with the idea that the changes in the SP-FMR signal comes from a direct REE contribution to the out-of-equilibrium spin density. To validate these results comprehensively, measurements of SP-FMR with varying DC current intensity were performed. Fig. 5(c) shows the peak value of SP-FMR for positive and negative values of the external applied currents. In all cases, a peak value of approximately -75 nA was observed. This result once again reinforces the idea of a contribution to $\vec{S}_{neq}$ due to the direct REE caused by the current passing through the sample. However, this extra component remains undetected due to the measurement geometry used.

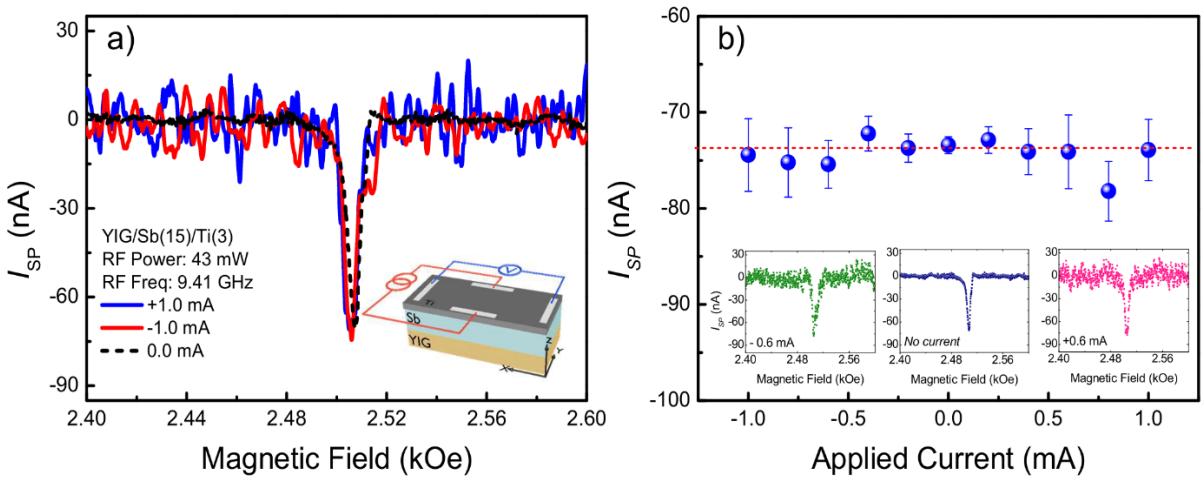

**Figure 5.** (a) A comparison of the spin pumping signal for an applied current of ±1.0 mA and no applied current, the insert is presented a schematic of the experimental setup used in the second part of the experiments carried out in this work. (b) The peak of the spin pumping signal in function of the applied current. In all cases investigated it was verified that the $I_{SP}$ signal remains constant.

- **CONCLUSIONS**

In conclusion, our findings reveal that Sb grown onto YIG indeed possess surface states, evidenced by the independence of the spin pumping signal on the injection direction. We further investigated the spin-to-charge conversion by introducing an external DC current into the system. Due to the spin-momentum locking, the non-equilibrium spin density produced in the experiment has two components: one originating from spin pumping and the other arising from the REE induced by the DC current. Thus, the measured signal is explained by:

$$\vec{J}_C = \frac{e\alpha_R}{\hbar}\left[\hat{z} \times \left(\vec{S}_{neq}^{SP} + \vec{S}_{neq}^{C}\right)\right]. \tag{4}$$

Surprisingly, our investigation demonstrates the ability to manipulate the polarity and intensity of the spin to charge conversion by varying the applied charge current. This capability paves the way for the development of low-power logic gates operating in the range of a few tens of microamperes.

**ACKNOWLEDGMENT**

This research is supported by Conselho Nacional de Desenvolvimento Científico e Tecnológico (CNPq), Coordenação de Aperfeiçoamento de Pessoal de Nível Superior (CAPES), Financiadora de Estudos e Projetos (FINEP), Fundação de Amparo à Ciência e Tecnologia do Estado de Pernambuco (FACEPE), Universidade Federal de Pernambuco, Multiuser Laboratory Facilities of DF-UFPE, Fundação de Amparo à Pesquisa do Estado de Minas Gerais (FAPEMIG) - Rede de Pesquisa em Materiais 2D and Rede de Nanomagnetismo, and INCT of Spintronics and Advanced Magnetic Nanostructures (INCT-SpinNanoMag), CNPq 406836/2022-1.

**ASSOCIATED CONTENT**

**Supporting Information**

 - Section I. The X-ray diffraction of the Sb films used in this work
 - Section II. Scanning tunneling microscopy (STM) and spectroscopy (STS) characterizations of Sb films.
 - Section III. Sample Fabrication
 - Section IV. Ferromagnetic Resonance (FMR) and Spin pumping measurements

**DATA AVAILABILITY STATEMENT**

The data that support the findings of this study are available from the corresponding authors upon reasonable request.

**NOTES**

The authors declare no competing financial interest.

## AUTHOR INFORMATION

**Corresponding Author**

*José E. Abrão - Email: elias_abrao@hotmail.com

*Joaquim B. S. Mendes - Email: joaquim.mendes@ufv.br

*Antonio Azevedo - Email: antonio.azevedo@ufpe.br

**Author Contributions**

Elias Abrao: Conceptualization (equal); Data curation (equal); Investigation (equal); Writing – original draft (equal). Eudes Gomes da Silva: Conceptualization (equal); Data curation (equal); Investigation (equal); Methodology (equal). Gilberto Rodrigues-Junior: Conceptualization (equal); Investigation (equal); Methodology (equal); Validation (equal); Writing – original draft (equal). Joaquim B. S. Mendes: Conceptualization (equal); Investigation (equal); Methodology (equal); Writing – original draft (equal); Writing – review & editing (equal). Antonio Azevedo: Conceptualization (equal); Formal analysis (equal); Funding acquisition (lead); Investigation (equal); Methodology (equal); Supervision (equal); Writing – original draft (equal); Writing – review & editing (equal).

# SUPPLEMENTARY INFORMATION

**Probing the Spin-Momentum Locking in Rashba Surfaces via Spin Current**

*José E. Abrão[1,*], Eudes Gomes da Silva[1,3], Gilberto Rodrigues-Junior[2], Joaquim B. S. Mendes[2,*], Antonio Azevedo[1,*]*

[1]Departamento de Física, Universidade Federal de Pernambuco, 50670-901 Recife, PE, Brazil.
[2]Departamento de Física, Universidade Federal de Viçosa, 36570-900 Viçosa, MG, Brazil.
[3]Department of Physics and Astronomy, University of Iowa, Iowa City, Iowa, USA.

**I. The X-ray diffraction of the Sb films used in this work**

The crystal structure and interface quality of sputtered Sb thin film on GGG (111) substrate were investigate by means of x-ray diffraction (XRD) and x-ray reflectivity (XRR) method using a four-circle high-resolution Bruker D8- discover diffractometer equipped with Cu $K_\alpha$ ($\lambda$ = 1.5418 Å) radiation source and a 2-bounce Ge (220) monochromator. Figure S1(a) shows a symmetrical (2θ-θ) XRD scan for the Sb thin film, performed in a high-resolution configuration (HRXRD). In addition to the peaks belonging the GGG substrate, it is possible to observe peaks that can be indexed with the Sb rhombohedral crystal structure with in-plane and out-of-plane lattice parameters of 4.30 Å and 11.22 Å, respectively.[1] Since in this configuration we are able to probe crystal planes that are oriented parallel to the substrate surface, the appearance of only (003n) diffraction peaks indicates a highly c-axis oriented growth of Sb on GGG (111) substrate.

In order to probe crystal planes that are not necessarily aligned to substrate surface as well as attenuate the diffracted signal associated with the GGG substrate, the overall crystalline structure of sputtered Sb film was characterized through x-ray diffraction in a grazing incidence configuration (GIDXRD). In the result presented in Figure S1(b) all diffraction peaks can be assigned to R-3m space group and point group $D_{3d}$, similar to expected for Sb powder x-ray diffraction pattern.[1] This feature corroborates the growth of high-quality Sb thin film without evidence of secondary phases formation

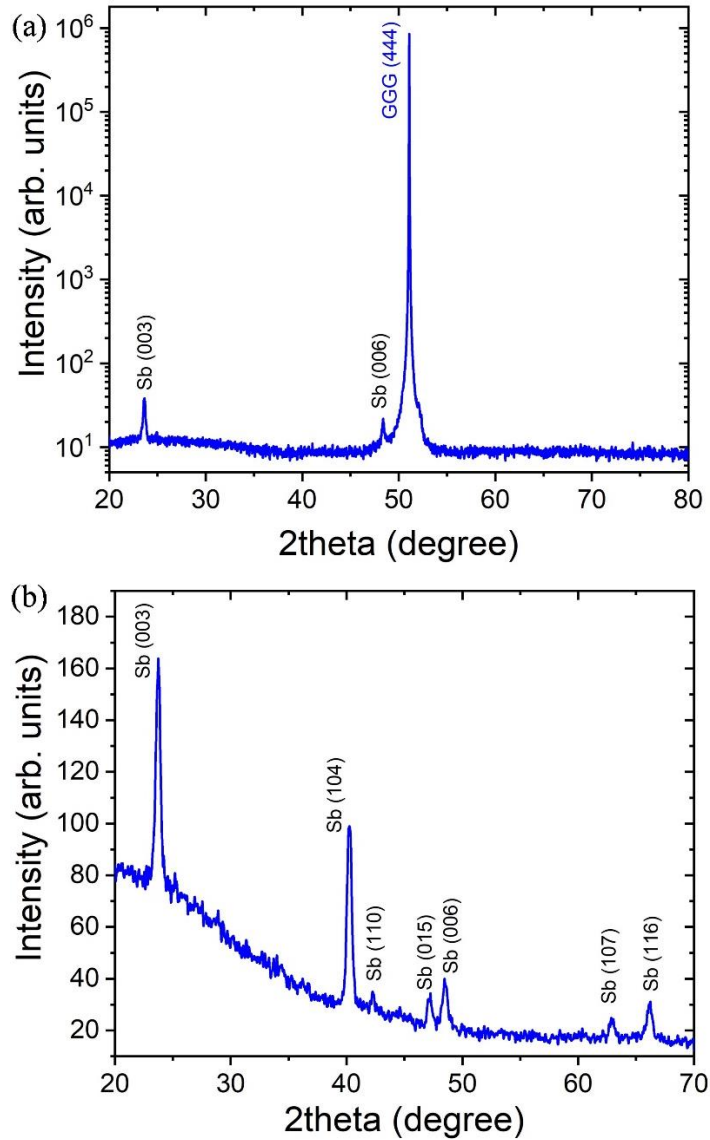

**Figure S1**. (a) Symmetrical HRXRD measurement and (b) Grazing incidence (GID) measurement of Sb thin film on GGG (111) substrate. All XRD peaks of Sb are indexed as a rhombohedral crystal structure with R-3m space group and point group $D_{3d}$.

To evaluate the film thickness and surface roughness, we carry out x-ray reflectivity measurements in the sputtered Sb thin films. The result is shown in Figure S2, expressed as a function of the longitudinal momentum transfer vector in the out-of-plane direction $q = 4\pi/\lambda \sin(2\theta/2)$, where $\lambda$ is the wavelength and $2\theta$ the scattering angle. The observation of an oscillatory pattern is direct evidence of a well-defined interface between Sb film and GGG substrate. The parameters obtained after XRR curve fitting (solid red line) are depicted in the graph. From these results, the total thickness of the film is 86.9 ± 0.5 nm and we can infer that the Sb thin films exhibit a low surface roughness (0.6 ± 0.2 nm) and density of 6.8 ± 0.2 g/cm³, in good accordance with the expected bulk value.

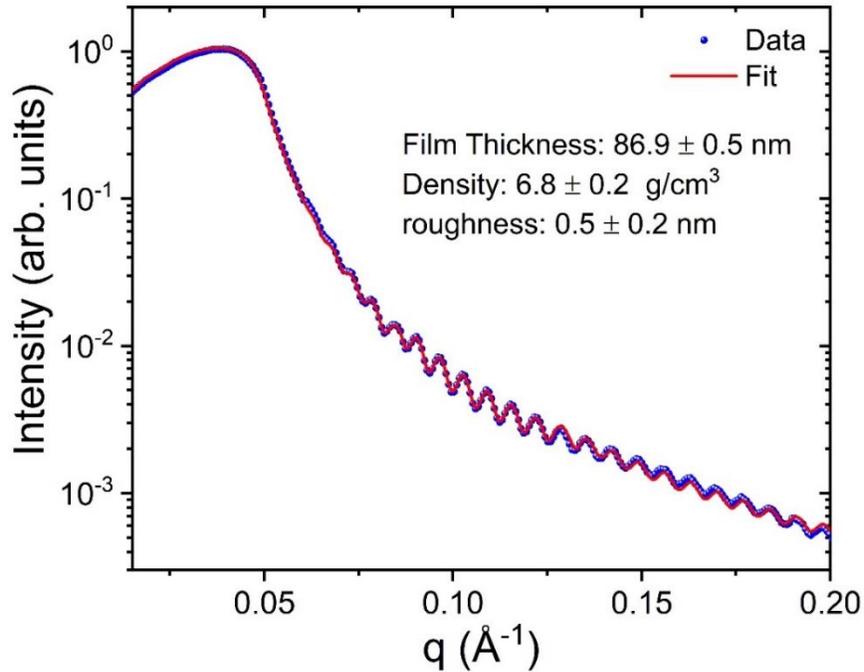

**Figure S2.** X-ray reflectivity curve (black dots) of Sb on GGG (111) substrate. The red solid line represents the curve fitting using the Parratt model. *Inset* shows the Sb thin film parameters obtained from fitting the experimental data.

## II. Scanning tunneling microscopy (STM) and spectroscopy (STS) characterizations of Sb films

Scanning tunneling microscopy (STM) and spectroscopy (STS) measurements were performed in order to investigate the surface electronic properties of Sb thin film. In a typical STS curve the differential tunneling conductance (dI/dV) is proportional to the local density of states (LDOS) at the STM tip position and, therefore, allows us to probe the presence of topologically protected states on the antimony surface.[2]

Figure S3 (left panel) shows a typical STM topographic image of Sb thin film evidencing a flat surface with root-mean-square (rms) roughness of about 0.8 nm in good agreement with our XRR results. Furthermore, the layered structure of sputtered Sb on GGG substrate can be observed due to the presence of well-defined terraces characterized by sharp steps corresponding to the inter Sb bilayer spacing with height of ~ 5.8 Å (Figure S3 right panel).[3]

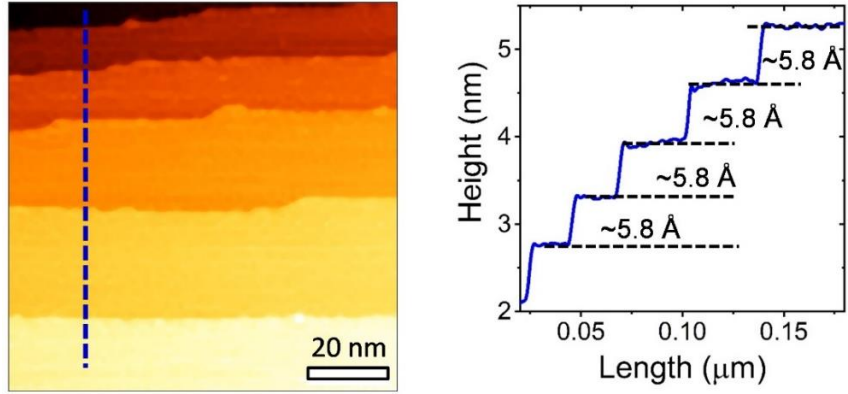

**Figure S3.** (left panel) STM topography image of about 100 × 100 nm$^2$ of the Sb surface at I = 0.3 nA and V = − 0. 5 V. (Right panel) Height profile taken along the blue dashed line indicated in the topography image.

Figure S4 shows a spatially averaged dI/dV spectrum acquired on Sb surface, the region of positive bias voltage denotes the bulk conduction band (BCB) and the region of negative bias voltage denotes the bulk valence band (BVB). The semimetal nature of Sb thin film can be observed through the suppression of LDOS in the vicinity of Fermi level (corresponding to zero bias voltage) indicating an overlapping between the conduction and valence bands.

In order to better understand the characteristics of the measured Sb STS spectrum it is necessary to analyze the overall antimony band structure, shown schematically as an inset in figure S4. The Sb semimetallic behavior results in a negative bulk band gap that distorts the topological surface states (TSS), resulting in a single pair of Rashba-split surface states that span the semimetallic gap with a Dirac cone at the zone center. [4,5] The inner cone can persist up to significantly higher energies, whereas the outer cone must fold down to connect with the bulk valence band. This leads to the appearance of extreme characteristics features within the outer Rashba cone, with one being a saddle point ($\varepsilon_S$) and the other a band edge ($\varepsilon_T$ and $\varepsilon_B$). [4-7]

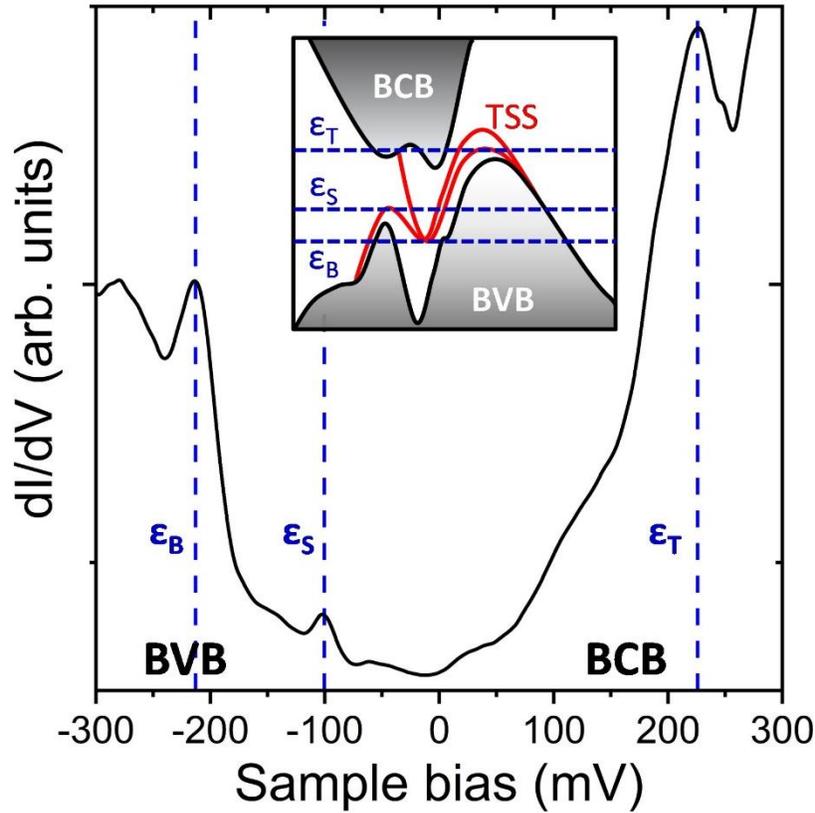

**Figure S4.** Spatially averaged dI/dV spectrum obtained along the Sb surface. BVB and BCB denotes the regions of bulk valence band and bulk conduction band, respectively. The vertical blue dashed line represents the $\varepsilon_B$, $\varepsilon_S$ and $\varepsilon_T$ topological surface states features. *Inset* shows a schematic representation of Sb electronic band structure (based on first principles calculations) representing the position of topological surface states with respect BVB and BCB. The zero energy (Sample bias) marks the Fermi level ($E_F$).

Based on this, the presence of three prominent peaks (vertical blue dashed lines) observed in dI/dV spectra shown in figure S4 can be a signature of TSS in sputtered Sb thin film. Indeed, the dI/dV peaks at − 214 mV and 227 mV are in good agreement with those observed in STS measurement in bulk antimony and can be associated with 2D band edges at $\varepsilon_B$ and $\varepsilon_T$, respectively.[4] Meanwhile, the dI/dV peak at − 104 mV usually is associated to the logarithmic singularity associated with the 2D saddle point at $\varepsilon_S$. [4-7]

**STM and STS Methods:** Scanning tunneling microscopy (STM) and spectroscopy (STS) measurements were performed under ultra-high vacuum conditions in an Omicron-VT STM system operating at room temperature, with a base pressure of $1,0 \times 10^{-10}$ mbar. All STM images were acquired using electrochemically etched polycrystalline tungsten (W) tips in constant current mode and for STS measurements a lock-in amplifier (operating a 3000 Hz) was used to obtain differential tunneling conductance (dI/dV) curves directly.

### III. Sample Fabrication

The Yttrium Iron Garnet (YIG - $Y_3Fe_5O_{12}$) stands out as a ferrimagnetic insulator with minimal magnetic loss, and which has been the reference material for exploring a variety of magnonic phenomena, including ferromagnetic resonance (FMR), the propagation of spin waves, optical transitions, the spin Hall effect, and the spin Seebeck effect.[8-17] YIG films were grown onto 1-inch diameter GGG ($Gd_3Ga_5O_{12}$) substrate via liquid phase epitaxy. $Fe_2O_3$, $Y_2O_3$, $B_2O_3$ and PbO powders were weighed and placed into a platinum crucible which was then inserted into a vertical tubular furnace, the furnace temperature was set to 940ºC with a rate of 1ºC/min. Once the furnace is up to temperature, the GGG substrate is then placed on a platinum sample holder that slowly descends into the crucible. To grow the YIG film, the GGG substrate dips into the melt inside the crucible, to ensure a uniform film the dipping process is done with the sample rotating at a fixed velocity.

Once the YIG is grown, the film is cut into 3x2 mm² samples with a low-speed diamond saw, following the samples are clean in an ultrasonic bath with acetone and isopropyl alcohol, the sample is then placed inside the sputtering chamber which was pumped down to a base pressure of $2.0 \times 10^{-7}$ torr or lower; The films were grown by DC sputtering in an argon atmosphere with a working pressure of $2.7 \times 10^{-3}$ torr. Figure S5 show a diagram of the sample fabrication process.

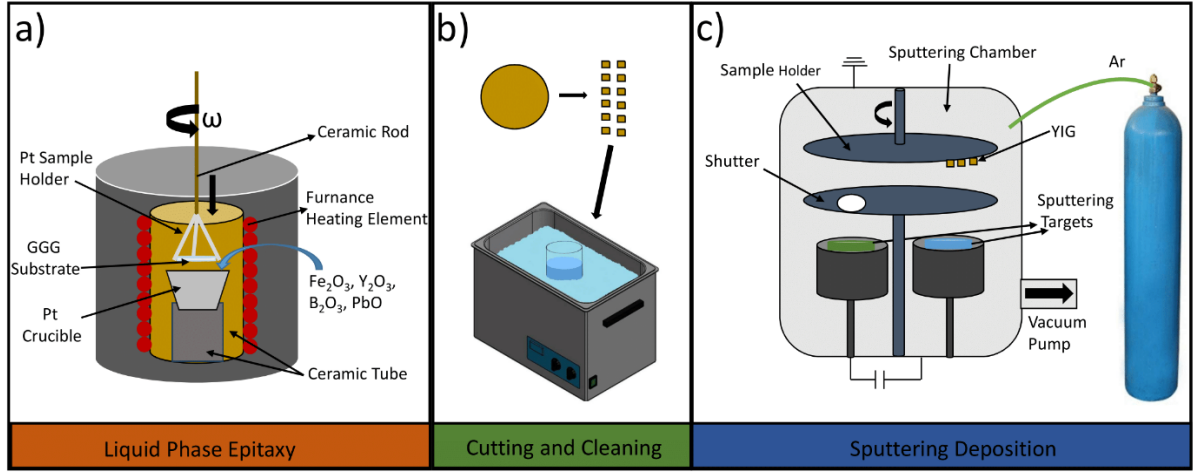

**Figure S5.** Schematics of the sample fabrication process, in (a) it is presented the liquid phase epitaxy method, in (b) it is show the cutting and cleaning procedures and in (c) it is shown the sputtering deposition process.

## IV. Ferromagnetic Resonance (FMR) and Spin pumping measurements

Ferromagnetic Resonance (FMR) measurements were performed in a home-made spectrometer. The sample is mounted on top of a PVC rod and through a small hole in a rectangular microwave cavity operating at $TE_{102}$, with resonance frequency of 9.42 GHz. The cavity is then placed inside the poles of an electromagnet, the microwave frequency used to excite the FMR condition is directed into the sample via a circulator and wave guides that are connected to the cavity. On the third side of the circulator a RF Schottky diode is placed which allows us to observe the reflected microwave power as function of the applied external field. To have a better noise-to-signal ratio the external field is modulated with two coils in the Helmholtz configuration which are supplied with an AC signal of 1.1 kHz, the diode signal is then send to a lock-in amplified. In a typical FMR experiment field sweeps are made while observing the reflected microwave power, at the resonance condition the sample will absolve the microwave radiation which will produces a drop in the signal that comes from the diode. Due to the use of a lock-in amplified, the observed signal ends up being the derivative of the reflected microwave power as function of the external field. There are two important properties that can be extract from a FMR specter, the resonance field $H_R$ and the linewidth $\Delta H$. The resonance field is related to the anisotropies of the material while the linewidth is related to dissipation mechanism of the film. Figure S6 shows a typical FMR spectra for YIG and YIG/Pt(2nm), where in Fig. S6 (a) we highlighted the two properties we can extract from the typical FMR spectra, note that by adding a thin layer of Pt the linewidth increases from 2.63 Oe to 2.84 Oe.

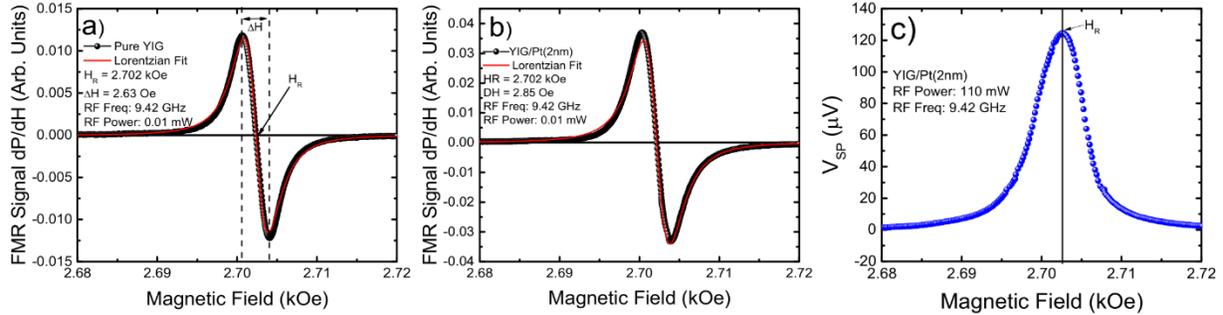

**Figure S6:** Typical FMR spectra for (a) pure YIG and (b) YIG/Pt(2nm). In (a) it is highlighted the Resonance field $H_R$ and the linewidth $\Delta H$, note that there is an increase in the linewidth because the addition of the Pt(2nm) layer. In (c) it is shown the spin pumping voltage measured by at the edges of the sample.

Spin pumping measurements were performed in the same spectrometer. The magnetization dynamics pumps a pure spin current from the ferromagnetic material into the adjacent layer, since the ferromagnetic material is losing energy in the form of a spin current being injected into the adjacent layer we observe an increase on the linewidth of the system. Inside the adjacent layer, the spin current can be converted into a charge current via inverse Rasbha effect or inverse Spin Hall effect. The electrical signal produced at the resonance condition can be detected via a nanovoltmeter by attaching two electrodes at the edges of the sample with silver paste. Figure S6(c) shows the spin pumping voltage obtained for YIG/Pt(2nm), note that the peak happens at the resonance field. The measured signal is normalized by the sample resistance to remove a geometric dependency of the spin pumping signal with the sample shape and size.